\documentclass[aps,pra,amsmath,amssymb,superscriptaddress,twocolumn,color,epsfig,graphicx,bm,floatfix]{revtex4-1}

\usepackage{amsmath}    % need for subequations
\usepackage{url}
\usepackage{hyperref}
\usepackage{mathtools}
\usepackage{siunitx}
\usepackage{multirow}
\usepackage{array}
\usepackage{tabulary}
\usepackage{longtable}
\usepackage[version=4]{mhchem}
\newcolumntype{K}[1]{>{\centering\arraybackslash}p{#1}}

\usepackage{graphicx}   % need for figures
\usepackage{verbatim}   % useful for program listings
\usepackage{color}      % use if color is used in text
\usepackage{subfigure}  % use for side-by-side figures
\usepackage{hyperref}   % use for hypertext links, including those to external documents and URLs
\usepackage{lipsum}
      
\begin{document}

\title{Modeling the atomtronic analog of an optical polarizing beam splitter, a half\textendash wave plate, and a quarter\textendash wave plate for phonons of the motional state of two trapped atoms}
\date{\today}
\pacs{???}
\author{Naeimeh Mohseni}
\email[]{n.mohseni@iasbs.ac.ir}
\affiliation{Department of Physics, Institute for Advanced Studies in Basic Sciences (IASBS), Iran}
\author{Marjan Fani}
\email[]{fani.m@sci.ui.ac.ir}
\affiliation{Department of Physics, University of Isfahan, Isfahan, Iran}
\author{Jonathan P. Dowling}
\email[]{jdowling@phys.lsu.edu}
\affiliation{Hearne Institute for Theoretical Physics and Department of Physics and Astronomy, Louisiana State University,  Baton Rouge, LA 70803 USA}
\author{Shahpoor Saeidian}
\email[]{saeidian@iasbs.ac.ir}
\affiliation{Department of Physics, Institute for Advanced Studies in Basic Sciences (IASBS), Iran}

\date{\today}
%----------------------------------------------------------------------------------------%
\begin{abstract}

In this paper we propose a scheme to model the phonon analog of optical elements, including a polarizing beam splitter, a half\textendash wave plate, and a quarter\textendash wave plate, as well as an implementation of {\footnotesize{CNOT} } and Pauli gates, by using two atoms confined in a two\textendash dimensional plane.  The internal states of the atoms are taken to be Rydberg circular states.  Using this model we can manipulate the motional state of the atom, with possible applications in optomechanical integrated circuits for  quantum information processing and quantum simulation. Towards this aim, we consider two trapped atoms  and let only one of them interact simultaneously  with two circularly polarized Laguerre\textendash Gaussian beams.
\end{abstract}

\maketitle
%*****************************************************************************************%
\section{Introduction\label{sec:introduction}}
  
Phonons can play a similar role to photons in quantum optics and quantum information processing. They can be used to encode information as qubits because of their appealing properties such as low\textendash propagation speed, which provides us with new schemes for processing quantum information, and their short wavelength, which allows us to access regimes of atomic physics that cannot be reached in photonic systems [\onlinecite{gustafsson2014propagating}]. Numerous researchers are trying to find ways of using phonons for quantum information and computation and,  more importantly, finding ways for manipulating the quantum information that is carried by these phonons [\onlinecite{soderberg2010phonon}, \onlinecite{zhu2006trapped}, \onlinecite{PhysRevB.88.064308}, \onlinecite{cirac1995quantum}, \onlinecite{toyoda2015hong}, \onlinecite{ruskov2012coherent}].

Cold and trapped atoms are good candidates that provide the possibility of using  phonons (vibrational motion of the trapped atoms) in quantum information and quantum optics. This system has attracted more attention because of its appealing properties such as long lifetimes (single atoms can  remain trapped for hours or days), long coherence times (ranging from milliseconds to seconds), natural reproducibility [\onlinecite{eltony2016technologies}], high controllability and large nonlinearity, which originate from quantization of the motion [\onlinecite{PhysRevA.75.042315}]. Moreover, integrated quantum atom chips have become the focus of current research in atomtronics, which promises the miniaturization (and therefore scaling) of optical quantum circuits [\onlinecite{leibrandt2009integrated}]. 
One important thing  is to show how we can manipulate motional states of the atoms, especially on quantum circuits. To achieve this aim, we have to implement arbitrarily quantum gates on motional states of the atoms.  On the other hand, as  Barenco, \textit{et al}., showed in 1995, any unitary operation can be decomposed into a sequence of  single\textendash qubit rotations and two\textendash qubit {\footnotesize{CNOT} } gates [\onlinecite{barenco1995elementary}]. Thus, by implementation of only these gates, we can realize arbitrary complex gates. The polarizing beam splitter (PBS), half\textendash wave plate (HWP), and quarter\textendash wave plate (QWP) perform operations of the two-qubit {\footnotesize{CNOT} } and single\textendash qubit  rotations gates. Wave plates and PBS operations on a trapped atom chip  have not been implemented before, but they are needed for the realization of universal quantum gates.  This paper is an effort in that regard.\\
  
 Here, we propose a scheme for modeling the phonon analog of optical elements including  the PBS, HWP, and QWP in such a way that the vibrational motion along the $ x\textendash $ and $ y\textendash $ axes plays the role of the horizontal and vertical polarization, respectively. To this end,  we consider two atoms trapped in the $x\textendash y$ plane, the internal states of the atoms of which are taken to be Rydberg circular states. Two counter\textendash propagating circularly polarized Laguerre\textendash Gaussian (LG) beams illuminate selectively only one of the atoms. We show that external degrees of freedom of the atoms can be decoupled from the internal degree of the first atom by preparing the initial internal state of the first atom in $ (\vert e\rangle +\vert g\rangle)/\sqrt{2} $ or $ (\vert e\rangle +i\vert g\rangle)/\sqrt{2} $, and adjusting the frequency of the beams and trap.  In this way we can model the PBS, HWP and QWP for the external degree of freedom of the atoms.
 % In this way the of the atoms does not affect our results. Thus, our model is quite robust against atomic spontaneous emission, which is important for practical implementation.
One  interesting feature of our scheme is the exploitation of orbital angular momentum (OAM) modes of light. Actually in our model the OAM modes lead to a nonlinearity, which is needed  for realizing the above\textendash mentioned phonon analog of optical elements.

In this paper we proceed as follows. In Sec. \ref{sec:Evolution operators}, we first determine evolution operators of polarization\textendash sensitive analog optical elements, including a PBS, a HWP, and a QWP. In Sec. \ref{sec:Realization}, we develop a method for the realization of the   PBS, HWP, and QWP for phonons. In other words, we engineer a Hamiltonian corresponding to the PBS, HWP, and QWP, by using two atoms trapped in a two\textendash dimensional plane, where one of them interacts with two classical, circularly polarized LG beams. In Sec. \ref{sec:gate}, we investigate the realization of Pauli  single\textendash qubit gates and a two\textendash qubit {\footnotesize{CNOT} } gate with the trapped atom. Finally, we summarize our results and conclude with Sec. \ref{sec:conclusion}.

\section{Evolution operators of polarizing optical elements\label{sec:Evolution operators}}
%In this section we review the evolution operators of polarization sensitive optical elements including PBS, HWP, and QWP.\\
\subsection{ Polarizing Beam Splitter}
\begin{figure}[h]
\centering
\includegraphics[width=0.22\textwidth]{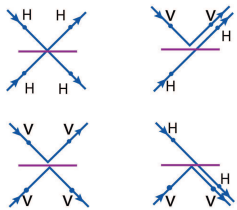}
\caption{ Schematic diagram of the PBS which  transmits the horizontal polarization and reflects the vertical polarization.
\label{fig:vgy1}}
\end{figure}

A PBS can be used for translation of spatial qubits into polarization qubits and is described by four degrees of freedom. Two of them are related to the spatial modes, and the other two are related to the polarization. In this optical element the horizontally polarized light is always transmitted, while the vertically polarized light is always reflected (Fig.\ref{fig:vgy1}.). Namely, the transmission coefficient for the horizontal mode $ t_{H} $  is 1 and for the vertical mode $ t_{V} $  is zero, and the reflection coefficient for the horizontal mode, $ r_{H} $ it is zero, and for the vertical mode $ r_{V} $ is 1.

\begin{figure}[h]
\centering
\includegraphics[width=0.22\textwidth]{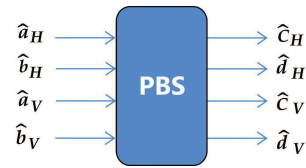}
\caption{A schematic diagram of operation of the PBS.
\label{fig:vgy22}}
\end{figure}

 So the operation of the PBS may be described by following matrix [\onlinecite{weihs20011}],
\begin{equation}
\begin{bmatrix}
   t_{H}   & ir_{H} & 0 & 0  \\
  i r_{H}  & t_{H} & 0 & 0 \\
   0   & 0 & t_{V} & i r_{V}\\
   0   & 0 & i r_{V} & t_{V}
\end{bmatrix}
%=
%\begin{bmatrix}
 %  1   & 0 & 0 & 0  \\
 %  0  & 1 & 0 & 0 \\
  % 0   & 0 & 0 & 1\\
   %0   & 0 & 1 & 0 
%\end{bmatrix}
,
\end{equation}
which shows that the PBS can be described by a matrix that is similar to the matrix of a {\footnotesize{CNOT} } gate[\onlinecite{weihs20011}]. Thus, mode transformations of a PBS may be described in the following form (Fig.\ref{fig:vgy22}), 
\begin{align}
&\hat{a}_{H}\rightarrow \hat{c}_{H}= \hat{a}_{H},\\
&\hat{b}_{H}\rightarrow \hat{d}_{H}= \hat{b}_{H},\\
&\hat{a}_{V}\rightarrow \hat{c}_{V}=i\hat{b}_{V},\\
&\hat{b}_{V}\rightarrow \hat{d}_{V}= i\hat{a}_{V},
\end{align} 
where $ \hat{a} $ and $  \hat{b}$ denote the input modes of the PBS and $ \hat{d} $ and $  \hat{c}$ denote the output modes of the PBS. In the following we will determine the unitary operator corresponding to the PBS. A general two-mode Hamiltonian, which can describe the creation of a photon in  mode $ a $ and the annihilation of a photon in  mode $ b $, and vice versa, may be written as [\onlinecite{kok2010introduction}], 
\begin{equation}
H_{\zeta, \varphi}=\hbar\zeta e^{i\varphi}\hat{a}^{\dagger} \hat{b}+\hbar\zeta e^{-i\varphi}\hat{a}\hat{b}^{\dagger},
\end{equation} 
%which describes the generalized beam splitter transformation. 
Under influence of this Hamiltonian the mode operators transformations are 
\begin{equation}
\begin{gathered}
e^{\frac{i}{\hbar}H t }\hat{a}e^{\frac{-i}{\hbar}H t }=\textrm{cos}(\zeta \textit{t} ) \hat{a}-ie^{i\varphi}\textrm{sin}(\zeta \textit{t}) \hat{b},\\
e^{\frac{i}{\hbar}H t}\hat{b}e^{\frac{-i}{\hbar}H t}=\textrm{cos}(\zeta \textit{t}) \hat{b}-ie^{-i\varphi}\textrm{sin}(\zeta \textit{t} ) \hat{a}.
\end{gathered}
\end{equation}
 %Then the unitary operator which gives rises to the polarizing beam splitter transformation will be as follows 
 If we set $ \zeta t=\pi /2 $ and $\varphi =0$, the unitary operator corresponding to this Hamiltonian will be,
\begin{equation}
 \begin{split}
 U=e^{ -i\frac{\pi}{2}(\hat{a}^{\dagger}\hat{b}+\hat{a}\hat{b}^{\dagger})},\label{1}
 \end{split}
 \end{equation} 
 which leads to the following transformations,
 \begin{eqnarray}
\hat{a}\rightarrow -i\hat{b},\\
\hat{b}\rightarrow -i\hat{a}.
\end{eqnarray}
 Therefore, the unitary operator corresponding to the PBS can be written as,
 \begin{equation}
 \begin{split}
 U=e^{ -i\frac{\pi}{2}(\hat{a}^{\dagger}_{V} \hat{b}_{V}+\hat{a}_{V}\hat{b}^{\dagger}_{V})}.\label{lkj}
 \end{split}
 \end{equation} 
This evolution operator leads to the following transformations,
 \begin{align}
 &\hat{a}_{H}\rightarrow \hat{a}_{H},\\
&\hat{b}_{H}  \rightarrow  \hat{b}_{H},\\
&\hat{a}_{V} \rightarrow -i\hat{b}_{V},\\
&\hat{b}_{V}  \rightarrow -i\hat{a}_{V},
 \end{align}
 which are similar to the PBS transformations regardless of %the factor of $-i$
 a minus sign in the third and forth transformations (which has no effect in our desired main result).  
\subsection{Wave Plates} 
 Another important optical instrument is the wave plate. A wave plate is an optical device that alters the polarization state of a light wave traveling through it. Two common types of wave plates are the HWP and QWP. For linearly polarized light, the HWP rotates the polarization vector through an angle $ 2\theta $, where $ \theta $ is the angle, which the optical axis of the material makes with the horizontal axis.  For elliptically polarized light, the HWP inverts the light's handedness [\onlinecite{kok2010introduction}]. The QWP converts linearly polarized light into circularly polarized light and vice versa. A QWP can be used to produce elliptical polarization as well.  In the following the unitary operators corresponding to these two optical elements are computed.\\
\subsubsection{Half\textendash Wave Plate}
\begin{figure}[h]
\centering
\includegraphics[width=0.32\textwidth]{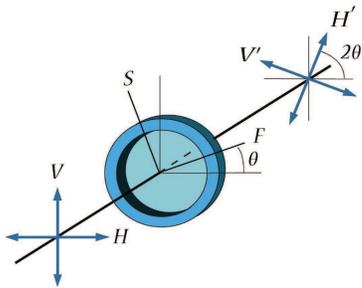}
\caption{ The operation of a HWP on a linearly polarized field. Axes parallel and perpendicular to the optical axis of the HWP are indicated by $F$ and $S$, respectively. A spatial rotation over an angle $ \theta$ of the wave plate induces a polarization rotation over an angle $ 2\theta $.\label{fig:vgy2}}
\end{figure}

 The mode transformations of a HWP have the following form [\onlinecite{kok2010introduction}],\\ 
\begin{eqnarray}
\hat{a}_{H}& \rightarrow & \textrm{cos}(2\theta)\hat{a}_{H}-i\textrm{sin}(2\theta)\hat{a}_{V},\\
\hat{a}_{V}&\rightarrow &-i \textrm{sin}(2\theta)\hat{a}_{H}+\textrm{cos}(2\theta)\hat{a}_{V}.
\end{eqnarray}
In the Bloch sphere, this corresponds to a rotation around the $ x\textendash$ axis. The unitary operator corresponding to the HWP transformation can be written as follows,
\begin{equation}
U= e^{ 2i\theta(\hat{a}^{\dagger}_{V} \hat{a}_{H}+\hat{a}_{V}\hat{a}^{\dagger}_{H})}.\label{mmm2}
\end{equation}
A rotation over an angle $  \theta$ of the HWP results in a polarization rotation over an angle $2\theta  $ (Fig.\ref{fig:vgy2}). 
%Therefore, a rotation over an angle $  \theta$ of the half-wave plate results in a polarization rotation over an angle $  2\theta$ .\\ 

\subsubsection{Quarter\textendash Wave Plate}
 The mode transformations of the QWP when the optical axis of the material is in the direction of the horizontal axis, have the following form,
\begin{eqnarray}
\hat{a}_{H}&\rightarrow & e^{-i\frac{\pi}{4}}\hat{a}_{H},\\
\hat{a}_{V} &\rightarrow & e^{+i\frac{\pi}{4}}\hat{a}_{V}.
\end{eqnarray}
Moreover, the Hamiltonian corresponding to the general single\textendash mode transformation may be written as [\onlinecite{kok2010introduction}],
\begin{equation}
\hat{H}=\hbar \varphi \hat{a}^{\dagger}\hat{a}. 
\end{equation}
This Hamiltonian leads to the following Bogoliubov transformation,
\begin{equation}
\hat{a}\rightarrow e^{-i\varphi}\hat{a}.
\end{equation}
 Thus the unitary operator, which describes a QWP is,
\begin{equation}
U=e^{i\frac{\pi}{4}[\hat{a}^{\dagger}_{V}\hat{a}_{V}-\hat{a}^{\dagger}_{H}\hat{a}_{H}]} .
 \end{equation}
In the following sections implementation of the phonon analog of the PBS, HWP, and QWP for the motional state of the two trapped atoms will be investigated.\\ 

\section{Realization of phonon analogs of PBS, HWP and, QWP for two trapped atoms\label{sec:Realization}}
Besides energy and linear momentum, photons carry spin\textendash  angular momentum (SAM) and orbital\textendash  angular momentum. SAM is associated with the polarization while OAM  is associated with the transverse amplitude and phase profile of the beam.  In this section, we use these degrees of freedom (OAM and SAM) to model phonon analogs of PBS, HWP, and QWP for the phonons.

\begin{figure}[h]
\centering
\includegraphics[width=0.22\textwidth]{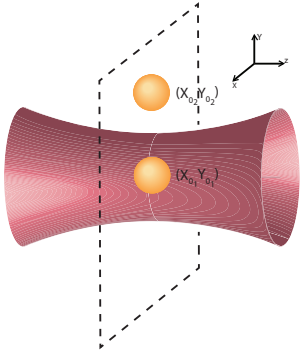}
\caption{ Interaction of two LG lights with one of the atoms confined in a transverse plane perpendicular to the beams at the waist of the beams.
\label{fig:bhu}}
\end{figure}

Let us consider two atoms trapped in an anisotropic three\textendash dimensional harmonic trap which is  described by the potential $U(\mathbf{r}_i)=1/2 m(\omega_{x_{i}}^2(x_i-x_{0_{i}})^2+\omega_{y_{i}}^2(y_i-y_{0_{i}})^2+\omega_z^2z_i^2)$,  where $m$ is the mass of the atoms and $ x_{0_{i}} $ and $ y_{0_{i}} $ are the minimum potential of the trapped atom $ i $. The $ \omega_{x_{i}} $, $ \omega_{y_{i}} $, and $ \omega_{z_{i}} $ are frequencies of the trap in the direction of $ x $, $ y $, and $ z $ for the trapped atom $ i $.  We assume that the atoms are tightly confined along the $z  $\textendash axis ($\omega_z\gg\omega_{x,y}$) and neglect the motion along this axis, and they oscillate  around ($ x_{0_{i}},y_{0_{i}} $) in the $ x\textendash y $ plane.  We can implement this kind of the trap by generalizing the  proposed trap in [\onlinecite{piotrowicz2013two}]. In this paper is introduced a blue detuned optical trap by using Gaussian beams propagating in direction of $ z $ and giving traps in the $ x\textendash y $ plane for Rydberg atoms.  Regarding the fact that blue detuned traps provide the possibility of simultaneous trapping of both ground and Rydberg excited states, they are interesting for experiments using Rydberg atoms.  We refer the reader to Ref. [\onlinecite{piotrowicz2013two}]for further details about how trapping fields interact with Rydberg excitation, temperature, and so on.\\
We introduce new coordinates as ($\hat{X}=(\hat{x}_{1}+\hat{x}_{2})/2$, $ \hat{Y}=(\hat{y}_{1}+\hat{y}_{2})/2$) and ($ \hat{x}=(\hat{x}_{2}-\hat{x}_{1})/2 $, $ \hat{y}=(\hat{y}_{2}-\hat{y}_{1})/2$), 
which shows the center of  mass (CM)  and  breathing mode operators, respectively.
Now consider two classical LG beams with the same amplitudes, polarizations (circular polarization), and transverse profile, but with different frequencies. Thus, the transverse electric field may be written as,
\begin{equation}
\textbf{E}=U_{\ell}(\rho,\phi)(\hat{e}_{x}+ i\hat{e}_{y})(e^{-i\omega_{1} t}+e^{-i\omega_{2} t})+\rm{h.c.},
\end{equation}
where $ U_{\ell}(\rho,\phi) $ is the transverse profile of LG beams which at the beam waist $ w_{0} $, is given by, 
 \begin{equation}
 U_{\ell}(\rho,\phi)=\varepsilon_{\ell}(\frac{\rho}{w_{0}})^{\vert \ell \vert } \textrm{exp}(-\frac{\rho^{2}}{w_{0}^{2}}+i\ell\phi),
 \end{equation}
 where $\rho$ and $ \phi $ are the radial and angular polar coordinates. The modes are characterized by an OAM equal to $ \ell \hbar $ along the propagation axis. We let only one of the atoms, say the first atom, to be illuminated simultaneously by the two classical lasers (Fig.\ref{fig:bhu}). This is possible experimentally, because first, we choose the potential of the trap in such a way that the equilibrium positions of the two atoms are spatially separated  and second, we assumed  the size of the CM mode wave function is small in comparison to the radius of the LG beams.  So if the laser forms an appropriate angle with the plane in which the atoms are trapped, it can illuminate selectively only one of the atoms  without illuminating the other one. For example this can be achieved if we apply a beam  with a waist  size  of  1.5 $ \mu \textrm{m} $
and a trap that can trap the atoms with spaces between atoms of the order of millimeters [\onlinecite{chu1998nobel}]. 

\begin{figure}[h]
\centering
\includegraphics[width=0.42\textwidth]{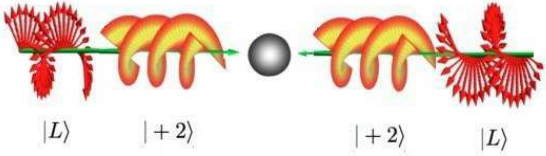}
\caption{ Interaction of two LG beams with the same amplitudes,  transverse profile,  and polarization (from the point of view of the atom) with an atom trapped in two dimensions.
\label{fig:bh1}}
\end{figure}

The  Hamiltonian of this system is given by,
\begin{equation}
\hat{H}=\hat{H}_{0}+\hat{H}_{\rm{int}}, \label{1-25}
\end{equation}%Let us consider the quantized circularly polarized Laguerre-Gaussian light propagating in the direction of $z$ that interact with a two level atom trapped in two-dimensions in presence a classical pump with frequency _$ \omega_{p} $ (in the $x-y$ plane perpendicular to the cavity quantized filed ). The internal levels in the atom are taken to be Rydberg circular states.
% moreover, this system is pumped by a classical circularly polarized Laguerre-Gaussian beam light at frequency of $ \omega_{p} $.
where

\begin{equation}
\begin{split}
\hat{H}_{0}&=  \frac{\hbar \omega_{0}}{2} \hat{\sigma}_{z_{1}}+ \hbar \nu_{x} \hat{a}^{\dagger}_{x} \hat{a}_{x}+\hbar \nu_{y} \hat{a}^{\dagger}_{y} \hat{a}_{y}\\
&  \quad +\hbar \mu_{x} \hat{b}^{\dagger}_{x} \hat{b}_{x}+\hbar\mu_{y} \hat{b}^{\dagger}_{y} \hat{b}_{y},  \label{1-26}
\end{split}
\end{equation}
and 
\begin{equation}
 \hat{H}_{\rm{int}}=-\hat{\textbf{P}}_{1}\cdot\textbf{E},
 \end{equation}
where $ \hat{\sigma}_{z_{i}} $ is the Pauli operator of the $i$th atom, $ \omega_{0} $ is the atomic frequency, $ \mu_{i}(\nu_{i}) $ is the frequency of the phonon in direction $ i =x,y $ for the  CM (breathing) mode, and $\hat{a}_{i}(\hat{b}_{i}) $ is the annihilation operator of the CM (breathing) mode in direction $ i $. $ \hat{\textbf{P}}_{1} $ is the dipole moment of the first atom which is given by
 \begin{equation}
 \hat{\textbf{P}}_{1}=\frac{1}{2}e[(\hat{x}_{1}+i\hat{y}_{1})(\mathbf{e}_{x}-i\mathbf{e}_{y})+(\hat{x}_{1}-i\hat{y}_{1})(\mathbf{e}_{x}+i\mathbf{e}_{y})],
 \end{equation}
where $  \mathbf{e}_{x}$ and $ \mathbf{e}_{y} $ are the unit vectors along $ x$ and $ y \textendash $ axis.
We assume the atoms to be in Rydberg circular electronic states.  Because of the fact that circular states have the maximum value of the magnetic quantum number, $ m=l=n-1 $, we represent $ \vert n=m+1,l=m,m\rangle $ with $ \vert m\rangle $ for simplicity and we denote two levels of the our atom with $ \vert m\rangle $ and $ \vert m+1 \rangle $. Thus, the dipole moment in the internal circular atomic basis can be written as,
 \begin{equation}
\begin{split}
 \hat{\textbf{P}}_{1}&=\frac{1}{2}P_{m}[(\mathbf{e}_{x}+i\mathbf{e}_{y})\hat{\sigma}_{m_{1}}+(\mathbf{e}_{x}-i\mathbf{e}_{y})\hat{\sigma}^{\dagger}_{m_{1}}]\\
& =\frac{1}{2}P_{m}[(\hat{\sigma}_{m_{1}}+\hat{\sigma}^{\dagger}_{m_{1}})\mathbf{e}_{x}+i(\hat{\sigma}_{m_{1}}-\hat{\sigma}^{\dagger}_{m_{1}})\mathbf{e}_{y}],
 \end{split}
 \end{equation}
where 
  \begin{equation}
e\langle m\vert(\hat{x}_{1}-i\hat{y}_{1})\vert m+1\rangle=e\langle m+1\vert (\hat{x}_{1}+i\hat{y}_{1})\vert  m\rangle=P_{m}.\\ 
 \end{equation} 
  and
 \begin{equation}
 \hat{\sigma}_{m_{1}}=\vert m\rangle \langle m+1\vert, \quad \hat{\sigma}^{\dagger}_{m_{1}}=\vert m+1\rangle \langle m\vert.
 \end{equation} 
Here, $ \hat{\sigma}_{m_{1}} $ and $ \hat{\sigma}^{\dagger}_{m_{1}} $ are Pauli operators of the first atom. \\
%In order to let the lasers illuminate selectively only one of the atoms, without illuminating the other one,   we can choose two  atoms with different transition frequencies,  but this may lead to a situation where the two atoms are no longer maintained in the trap.
%, because  for trapping, the frequency of the trap beam have to  be blue detuned from the atomic transition.
 %To solve this problem, we assume that the atoms have three levels, in such a way that the transition frequency of two levels (say the ground state $|g\rangle$ and the first excited state $|e_1\rangle$) is the same for the two atoms, while the other transition frequencies of the two atoms are different ( Fig.\ref{fig4}).  For trapping we use the levels with same transition frequencies and for interaction of the first atom with the  two LG beams we can exploit another set of two levels.
 %According to what is mentioned in the previous section
%\begin{figure}[h]
%\centering
%\includegraphics[width=0.35\textwidth]{5.pdf}
%\caption{(color online) Atomic transition frequencies.
%\label{fig4}}
%\end{figure}
By assuming that the size of the CM mode wave function $R_0$ is small compared with the radius  of the LG beams $w_0$, and given the fact that the CM motion is quantized,  the transverse profile of the electric field at the beam waist and in the place of the first atom, for $ \ell\in 0$,  $\pm 1$,  $\pm2, ..., $ can be written as [\onlinecite{1464-4266-4-2-371}],
  \begin{equation}
\begin{split}
 U_{\ell}(\hat{\rho},\hat{\phi})&=\varepsilon_{\ell}(\frac{\hat{\rho}}{w_{0}})^{\vert \ell \vert} \textrm{exp}(il\hat{\phi})=\frac{\varepsilon_{\ell}}{w_{0}^{\vert \ell\vert}} ( \hat{x}_{1}\pm i\hat{y}_{1})^{\vert \ell \vert}\\
 &=\frac{\varepsilon_{\ell}}{w_{0}^{\vert \ell \vert}}((\hat{X}-\hat{x})\pm i(\hat{Y}-\hat{y}))^{\vert \ell \vert}\\
 &=\varepsilon_{\ell}[(\eta_{x_{c}}(\hat{a}^{\dagger}_{x} +\hat{a}_{x})-\eta_{x_{b}}(\hat{b}^{\dagger}_{x} +\hat{b}_{x}))\\
 &\quad \pm i(\eta_{y_{c}}(\hat{a}^{\dagger}_{y} +\hat{a}_{y})-\eta_{y_{b}}(\hat{b}^{\dagger}_{y} +\hat{b}_{y}))]^{\vert \ell\vert}
 \end{split}
 \end{equation}
  where $ \eta_{x_{c}}=\sqrt{\hbar/m\mu_{x}w_{0}^{2}}$, $ \eta_{y_{c}}=\sqrt{\hbar/m\mu_{y}w_{0}^{2}}$, $ \eta_{x_{b}}=\sqrt{\hbar/m\nu_{x}w_{0}^{2}}$, $ \eta_{y_{b}}=\sqrt{\hbar/m\nu_{y}w_{0}^{2}}$.\\ 
The interaction Hamiltonian of this system, for $ \ell\in 0$, $\pm 1$,  $\pm2,..., $ is given by,

\begin{equation}
\begin{split}
\hat{H}_{\rm{int}}= -\frac{\hbar}{2} \Omega_{m,\ell}[(\eta_{x_{c}}(\hat{a}^{\dagger}_{x} +\hat{a}_{x})-\eta_{x_{b}}(\hat{b}^{\dagger}_{x} +\hat{b}_{x}))\\
\pm i(\eta_{y_{c}}(\hat{a}^{\dagger}_{y} +\hat{a}_{y})-\eta_{y_{b}}(\hat{b}^{\dagger}_{y} +\hat{b}_{y}))]^{\vert \ell\vert}\\
\times \hat{\sigma}^{\dagger}_{m_{1}}(e^{-i\omega_{1} t}+e^{-i\omega_{2} t})+\rm{h.c.} , 
 \label{1-27}
\end{split} 
\end{equation}
where $ \Omega_{m,\ell}= 2P_{m} \varepsilon_{\ell}/\hbar $,  which is assumed to be equal for the two standing\textendash wave lasers. We select two lasers with the following frequencies, 
\begin{equation}
\omega_{1}=\nu_{y} -\mu_{y}+\omega_{0}, \quad \omega_{2}=-(\nu_{y} -\mu_{y})+\omega_{0}.
\end{equation}
Under this condition, the Hamiltonian in the interaction picture for $ \ell=2 $ (Fig.\ref{fig:bh1}), in the rotating\textendash wave approximation, will be,
\begin{equation}
\hat{H}_{\rm{int}}=\mp \hbar \Omega_{m,2} \eta_{y_{c}} \eta_{y_{b}}(\hat{a}^{\dagger}_{y}\hat{b}_{y}+\hat{a}_{y}\hat{b}^{\dagger}_{y})(\hat{\sigma}^{\dagger}_{m_{1}}+\hat{\sigma}_{m_{1}}).\label{ghu}
\end{equation}
 The evolution operator corresponding to this Hamiltonian is,
\begin{equation}
U=\textrm{exp}[\pm i\Omega_{m,2} \eta_{y_{c}} \eta_{y_{b}}t(\hat{a}^{\dagger}_{y}\hat{b}_{y}+\hat{a}_{y}\hat{b}^{\dagger}_{y})(\hat{\sigma}^{\dagger}_{m_{1}}+\hat{\sigma}_{m_{1}})]. 
\end{equation}
By preparing the initial internal state of the first atom in $ (\vert e\rangle +\vert g\rangle)/\sqrt{2} $,  the internal and external degrees of freedoms would be decoupled. So dynamical evolution of the motional states of the atoms is decoupled from the dynamical evolution of internal states of the first atom.  Moreover, the motional part of the evolution operator is similar to the evolution operator of the PBS, if we set $ \Omega_{m,2} \eta_{y_c} \eta_{y_b}t=\pi/2 $. 
%Moreover, by preparing the initial internal state of the first atom in $ (\vert e\rangle +\vert g\rangle)/\sqrt{2} $  the internal and external degrees of freedoms would be decoupled.
 %, so the spontaneous emission does not affect the dynamical evolution of the external degrees of freedom. Namely, our modeling  is quite robust against atomic spontaneous emission. \\ 
This condition can be achieved and is compatible with experimental values found in the literature [\onlinecite{hogan2012cold},\onlinecite{semiao2002nonclassical}]. In more detail, the internal levels of the atoms are taken to be Rydberg circular states. These states are extremely long lived, with lifetimes that scale as $n^{5}  $,  which are of the order of $ 10^{-2}$  $ \rm{s} $ for $ n = 30 $, and the electric dipole moments are of the order of $ 10^{-27}$   $ \rm{Cm} $ [\onlinecite{hogan2012cold}]. The Lamb-Dicke parameter can be of the order of $10^{-1}$ [\onlinecite{semiao2002nonclassical}]. Therefore, a laser with an intensity of $ 1$   $\rm{ W m^{2}} $, corresponding to a power of $ 10^{-6} $   $\rm{W}$,  can satisfy  the condition $ \Omega_{m,2} \eta_{y_c} \eta_{y_b}t=\pi/2 $ for an interaction time $ 10^{-5}$   $\rm{s} $. This shows that, not only the condition $\Omega_{m,2} \eta_{y_c}\eta_{y_b}t=\pi/2 $ can be achieved, but also the interaction can be carried out before spontaneous emission can occur.  
  
 %Thus we can say our system under mentioned conditions act as a PBS.
It is worth noting that the  operators $ \hat{a}_{x} $ and $ \hat{a}_{y} $ play the analog role of $\hat{a}_{H} $ and $ \hat{a}_{V} $, respectively. In other words, the phonons  vibrating in the direction of the $ x  $ ($  y$) axis play the role of the horizontal (vertical) component of polarization.
% Actually this is only a modeling, 
In more detail, an optical PBS reflects vertical photons and  transmits horizontal photons,  while under the effect of our analog PBS, the phonons of the  CM mode and breathing mode vibrating in the direction of the $ x \textendash$ axis  will remain  in the same mode, but the  phonons of the breathing mode vibrating in the direction of the $ y\textendash $ axis will change to the CM mode phonons vibrating in the direction of the $ y \textendash $ axis and vice versa.\\

In an other case, if we select two lasers with the following frequencies,
\begin{equation}
\omega_{1}=\mu_{x} -\mu_{y}+\omega_{0} \quad \omega_{2}=-(\mu_{x} -\mu_{y})+\omega_{0},
\end{equation}
 the Hamiltonian in the interaction picture for $ \ell=2 $ will be,
\begin{equation}
\hat{H}_{\rm{int}}=\mp i\hbar \Omega_{m,2} \eta_{y_{c}} \eta_{x_{c}}(\hat{a}^{\dagger}_{x}\hat{a}_{y}+\hat{a}_{x}\hat{a}^{\dagger}_{y})(\hat{\sigma}^{\dagger}_{m_{1}}-\hat{\sigma}_{m_{1}}).
\end{equation}
The evolution operator corresponding to this Hamiltonian is,
\begin{equation}
U=\textrm{exp}[ \pm\Omega_{m,2} \eta_{x_{c}} \eta_{y_{c}}t(\hat{a}^{\dagger}_{x}\hat{a}_{y}+\hat{a}_{x}\hat{a}^{\dagger}_{y})(\hat{\sigma}^{\dagger}_{m_{1}}-\hat{\sigma}_{m_{1}})].
\end{equation}
In this case, by preparing the initial internal state of the first atom in $ (\vert e\rangle +i\vert g\rangle)/\sqrt{2} $,  the internal and external degrees of freedoms would be decoupled. So dynamical evolution of the motional states of the atoms and dynamical evolution of the internal states of the first atom are decoupled.  Moreover, the motional part of this evolution operator is similar to the evolution operator of a HWP; i.e.,  this interaction acts as a HWP for CM phonons. Actually this HWP rotates the direction of the vibration of the CM phonon, by $ 2\theta $, where, $ \theta = \Omega_{m,2}\eta_{x_c}\eta_{y_c}t/2 $, and can be controlled by adjusting the Lamb-Dicke parameters $ \eta_{x_c}$ and  $\eta_{y_c} $, the coupling strength of the atom-light interaction, and the time of the interaction. These features provide good possibilities for manipulating the vibrational states of the atoms, which can find application in quantum information processing and computing.\\% For this case too, if we prepare the initial internal state of the first atom in $ (\vert e\rangle +\vert g\rangle)/\sqrt{2} $, dynamical evolution of the internal state would be decoupled from the external degrees of freedom.\\
 If we select two lasers with the following frequencies, 
\begin{equation}
\omega_{1}=\nu_{x} -\nu_{y}+\omega_{0} \quad \omega_{2}=-(\nu_{x} -\nu_{y})+\omega_{0},
\end{equation}
The interaction will act as a HWP for the breathing phonons.\\
In the case that  we select two lasers with the following frequencies, 
\begin{equation}
\omega_{1}=\omega_{2}\simeq \omega_{0},
\end{equation}
and suppose $ \mu_{i}  \ll \nu_{i}$, the interaction Hamiltonian will be, 
\begin{equation} 
\hat{H}_{\rm{int}}= -\hbar\Omega_{m,2}[\eta_{x_c}^{2}\hat{a}^{\dagger}_{x}\hat{a}_{x}-\eta_{y_c}^{2}\hat{a}^{\dagger}_{y}\hat{a}_{y}+(\eta_{x_c}^{2}-\eta_{y_c}^{2})](\hat{\sigma}^{\dagger}_{m_{1}}+\hat{\sigma}_{m_{1}}).
 \label{abc}
\end{equation}
We can rewrite this Hamiltonian as,
\begin{equation}
\begin{split}
\hat{H}_{\rm{int}}&=-\hbar\Omega_{m,2} [\frac{\eta_{x_c}^{2}-\eta_{y_c}^{2}}{2}(\hat{a}^{\dagger}_{x}\hat{a}_{x}+\hat{a}^{\dagger}_{y}\hat{a}_{y})\\
& \quad+\frac{\eta_{x_c}^{2}+\eta_{y_c}^{2}}{2}(\hat{a}^{\dagger}_{x}\hat{a}_{x}-\hat{a}^{\dagger}_{y}\hat{a}_{y})
+(\eta_{x_c}^{2}-\eta_{y_c}^{2})] (\hat{\sigma}^{\dagger}_{m_{1}}+\hat{\sigma}_{m_{1}}).\label{abcer1}
\end{split}
\end{equation}
The evolution operator corresponding to this Hamiltonian is,
\begin{equation} 
\begin{split}
U &=  \textrm{exp}[(\eta_{x_{c}}^{2}-\eta_{y_{c}}^{2})(\hat{\sigma}^{\dagger}_{m_{1}}+\hat{\sigma}_{m_{1}})]\\
 & \quad  \times \textrm{exp}[i\Omega_{m,2}\frac{\eta_{x_{c}}^{2}-\eta_{y_{c}}^{2}}{2}t(\hat{a}^{\dagger}_{x}\hat{a}_{x}+\hat{a}^{\dagger}_{y}\hat{a}_{y})(\hat{\sigma}^{\dagger}_{m_{1}}+\hat{\sigma}_{m_{1}})] \\ 
& \quad \quad \times \textrm{exp}[i\Omega_{m,2}\frac{\eta_{x_{c}}^{2}+\eta_{y_{c}}^{2}}{2}t(\hat{a}^{\dagger}_{x}\hat{a}_{x}-\hat{a}^{\dagger}_{y}\hat{a}_{y})(\hat{\sigma}^{\dagger}_{m_{1}}+\hat{\sigma}_{m_{1}})]. \label{wer}
 \end{split}
\end{equation}
As we see, if we prepare conditions under which $ \Omega_{m,2}(\eta_{x_c}^{2}+\eta_{y_c}^{2})t/2=\pi/4 $  (this can be achieved corresponding to experimental values that we mentioned earlier), and we prepare the initial internal state of the first atom  prepare in $ (\vert e\rangle +\vert g\rangle)/\sqrt{2} $, the system acts as a QWP on the CM mode, irrespective of the phase of $ \textrm{exp}[i\Omega_{m,2}(\eta_{x_c}^{2}-\eta_{y_c}^{2})(\hat{a}^{\dagger}_{x}\hat{a}_{x}+\hat{a}^{\dagger}_{y}\hat{a}_{y})t/2] $. (Note that the number of phonons, $N=\hat{a}^{\dagger}_{x}\hat{a}_{x}+\hat{a}^{\dagger}_{y}\hat{a}_{y}$ is a constant of motion.) 
If $ \mu_{i} \gg \nu_{i}$  and $ \Omega_{m,2}(\eta_{x_b}^{2}+\eta_{y_b}^{2})t/2=\pi/4 $ the interaction acts as a QWP on the breathing phonons.\\
Therefore, only by adjusting the frequencies of the two LG beams (with $ \ell=2 $), and the trap, one can prepare conditions such that the interaction acts as a PBS, HWP, and QWP for vibrational phonons.
%*****************************************************************************************
\section{Realization of CNOT and Pauli gates\label{sec:gate}}
\begin{figure}[h]
\centering
\includegraphics[width=0.42\textwidth]{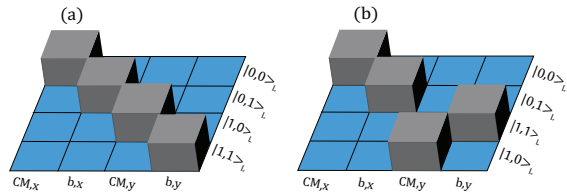}
\caption{ \textrm {(a)} ‍Characterization of the two-qubit logical states. \textrm{(b)} The effect of the PBS interaction (36) on the input two-qubit states. 
\label{fig:bhnm}}
\end{figure}

%One of the requirements for the physical implementation of quantum computing in a certain quantum system is a set of quantum gates that can be realized in the quantum system under consideration.
As Barenco \textit{et al}., showed, any unitary operation can be decomposed into a sequence of single\textendash qubit rotations and two\textendash qubit {\footnotesize{CNOT} } gates [\onlinecite{barenco1995elementary}]. In this section we show how one can realize quantum  {\footnotesize{CNOT} } and Pauli gates by using the interactions that we have investigated here.\\
If we encode vibrating directions ($x$ and $y$) of phonons as states of the control qubit ($|0\rangle_c=|n_x=1, n_y=0\rangle$ and $|1\rangle_c=|n_x=0, n_y=1\rangle$) and the type of the phonon mode (CM or breathing) as states of the target qubit ($|0\rangle_t=|n_{\rm{CM}}=1, n_b=0\rangle$ and $|1\rangle_t=|n_{\textrm{CM}}=0, n_b=1\rangle$), the four two-qubit input states ($|0,0\rangle_{L}=|n_x^c=1, n_x^b=0, n_y^c=0, n_y^b=0\rangle$, $|0,1\rangle_{L}=|0, 1, 0, 0\rangle$, $|1,0\rangle_{L}=|0, 0, 1, 0\rangle$, $|1,1\rangle_{L}=|0, 0, 0, 1\rangle$) can be generated. In this case the PBS interaction Eq. (\ref{ghu}) acts as a {\footnotesize{CNOT} } gate on these qubits (Fig. \ref{fig:bhnm}). In the same way, single\textendash qubit gates such as the  Pauli  $X$ gate and Pauli  $Z$ gate can be realized by using HWP and QWP interactions.\\
%if we encode phonons vibrate in direction of $ x $ ($ y $) as qubit $ \vert 0\rangle $ ( $ \vert 1\rangle $).
Now, let us consider the preparing of the initial states and readout of the final state. To prepare the initial state, we should cool down the atoms to be prepared in their motional ground states, then by illuminating a sequence of pulses with desired frequencies on the atoms we can excite the motional state of them in the breathing motion or CM motion in each of the $  x$ and $  y$ directions. Regarding the readout process, the motional state of the trapped atom could be measured in two steps: At first, the external state of the atom should map into the internal state by using a Jaynes-Cummings interaction, and then by performing a measurement on the internal states, one can determine the probability distribution of the motional state [\onlinecite{1464-4266-6-7-002}].
It should be noted that the decoherence processes cannot restrict the operation of the above mentioned gate, considerably. The most important decoherence effects in our system are spontaneous emission of atoms, mechanical damping, and long-range Rydberg interaction. As we mentioned before, the time scales of operations of the {\footnotesize{CNOT} } and Pauli gates are smaller than the lifetime of the considered Rydberg atoms. Thus, we can be sure that during the interaction the spontaneous emission does not occur. Moreover, mechanical damping is of the order of several seconds, but we have estimated the duration of our scheme to be of the order of microseconds, so the mechanical damping can be neglected, too. On the other hand, long-range dipole-dipole interaction between the two Rydberg atoms can also be neglected, because it is proportional to  $ (1/r)^{6} $ and $ P^{4} $, where $ r $ and $ P $ are the distance between the atoms and dipole moment of the atoms, respectively. The dipole moment of the Rydberg atoms is proportional to  $  a_{0}n^{2} $, in  which $  a_{0}$ is the Bohr radius and $ n $ is the principal quantum number. For our case the distance between the atoms is of the order of millimeters and $ n=30 $, so dipole-dipole  interaction is of the order of $  10^{-1}\textrm{Hz}$, which when compared  to coupling strength of the atom with light ($ \Omega\sim 10^{7} \textrm{HZ}$) is negligible [\onlinecite{casimir1948influence}].

\section{Conclusion\label{sec:conclusion}}
In this paper we propose a scheme  for modeling the phonon analog of the optical elements including the PBS,  HWP, and QWP, as well as an implementation of {\footnotesize{CNOT} } and Pauli gates, by using two trapped atoms,  one of which interacts with two circularly polarized LG beams. This implementation can find application in the manipulation of quantum states of the phonons for realization of quantum information and quantum computing goals in integrated atom-optical circuits.\\

\begin{acknowledgments}
JPD would like to acknowledge support from the U. S. National  Science Foundation. This paper was supported by the Institute for Advanced Studies in Basic Sciences (Grant No.G2015IASBS12648).
\end{acknowledgments}
%*****************************************************************************************
%\appendix
%\section{}
%*****************************************************************************************
%\clearpa
\bibliography{Mohseni}% Produces the bibliography via BibTeX.

\end{document}